\documentclass[a4paper]{article}
\usepackage[english]{babel}
\usepackage[T2A]{fontenc}
\usepackage[utf8]{inputenc}
\usepackage{amsmath}
\usepackage{amsthm}
\usepackage{amssymb}
\usepackage{textcomp}
\usepackage{graphicx}
\usepackage[unicode,pdfborder={0 0 1}]{hyperref}

\usepackage{todonotes}

\newcommand{\eqlb}[2]{\begin{equation} \label{#1} #2 \end{equation}}
\newcommand{\eq}[1]{\begin{equation} #1 \end{equation}}
\newcommand{\eqn}[1]{\begin{equation}\nonumber #1 \end{equation}}

\newcommand{\eqs}[1]{$#1$}
\newcommand{\brc}[1]{\left(#1\right)}
\newcommand{\bsq}[1]{\left[#1\right]}
\newcommand{\bfi}[1]{\left\{ #1\right\}}

\newcommand{\abs}[1]{\left|#1\right|}

\newcommand{\qq}{\qquad}
\newcommand{\ds}{\displaystyle}

\newcommand{\wt}[1]{\widetilde{#1}}
\newcommand{\matr}[2]{\begin{array}{#1}#2\end{array}}

\newcommand{\at}[2]{\genfrac{}{}{0pt}{}{#1}{#2}}

\newcommand{\rme}{\textrm{e}}

\numberwithin{equation}{section}

\hoffset= - 1.0in
\def\Jac{\mathop{\rm Jac}\nolimits}

\begin{document}

\title{{\bf {Seiberg-Witten curves  and double-elliptic integrable systems} \vspace{.2cm}}
\author{{\bf G. Aminov$^{a,b,}$}\footnote{aminov@itep.ru}, {\bf H.W. Braden$^{c,}$}\footnote{hwb@ed.ac.uk}, {\bf A. Mironov$^{d,a,e,}$}\footnote{mironov@itep.ru; mironov@lpi.ru}, {\bf A. Morozov$^{a,e,}$}\thanks{morozov@itep.ru} \ and {\bf A. Zotov$^{f,a,b,}$}\thanks{zotov@mi.ras.ru; zotov@itep.ru}}
\date{ }
}

\maketitle

\vspace{-6.0cm}

\begin{center}
\hfill FIAN/TD-13/14\\
\hfill ITEP/TH-30/14\\
\end{center}

\vspace{4.2cm}

\begin{center}
$^a$ {\small {\it ITEP, Moscow 117218, Russia}}\\
$^b$ {\small {\it Moscow Institute of Physics and Technology, Dolgoprudny 141700, Russia }}\\
$^c$ {\small {\it School of Mathematics,
University of Edinburgh, Edinburgh, Scotland}}\\
$^d$ {\small {\it Lebedev Physics Institute, Moscow 119991, Russia}}\\
$^e$ {\small {\it Moscow Physical Engineering Institute, Moscow 115409, Russia }}\\
$^f$ {\small {\it Steklov Mathematical Institute, RAS, Moscow, Russia}}
\end{center}

\vspace{1cm}

\begin{abstract}
An old conjecture claims that commuting Hamiltonians
of the double-elliptic integrable system are constructed
from the theta-functions associated with Riemann
surfaces from the Seiberg-Witten family,
with moduli treated as dynamical variables
and the Seiberg-Witten differential providing the pre-symplectic
structure.
We describe a number of theta-constant equations
needed to prove this conjecture for the $N$-particle
system. These equations provide an alternative method to derive
the Seiberg-Witten prepotential and we illustrate this by calculating the perturbative contribution.
We provide evidence that the
solutions to the commutativity equations are exhausted by the double-elliptic system and its
degenerations (Calogero and Ruijsenaars systems).
Further, the theta-function identities that lie behind the Poisson commutativity of the three-particle Hamiltonians are proven.
\end{abstract}

\section{Intoduction}

The discovery of Seiberg-Witten theory \cite{SW1,SW2} and related integrable systems \cite{GKMMM} in the  mid-nineties  gave rise to many new insights in the theory of integrable systems. In particular it led to a new understanding \cite{BMMM'2000,FGNR,MM,MM2,GM,BGOR,AMMZ} of the classical $p-q$ duality \cite{Ruj'87,Ruijs} of the Calogero-Ruijsenaars family \cite{GM} via its relation with the Seiberg-Witten construction of the low-energy limit of ${\cal N}=2$ SUSY gauge theories with adjoint matter hypermultiplets: the $4d$ theory is described by the elliptic Calogero-Moser system (the elliptic modulus being associated in physical theory with the bare coupling constant) \cite{DW,IM'1996,IM'1997}, the $5d$ theory (with one compactified Kaluza-Klein dimension) is described by the elliptic Ruijsenaars system \cite{BMMM'1999} and the $6d$ theory (with two compactified Kaluza-Klein dimensions) is described by the double-elliptic integrable system (the second elliptic modulus being associated with the compactification torus or an abelian surface) \cite{BMMM'2000,MM,MM2,Braden2003}. All these integrable systems have an elliptic dependence on particle momenta and the most interesting ones are the double-elliptic integrable systems, where both coordinates and momenta have compact values.

The $p-q$ duality admits various deformations. For instance, with a one parametric deformation one can lift it to quantum integrable systems \cite{Ruj'87,Ruijs,EV,FMTV,TV,GM} where the $p-q$ duality can be studied at the level of wave functions; this provides an additional tool to work with. This deformation corresponds to the Nekrasov-Shatashvili limit, $\epsilon_2\to 0$ \cite{NS2009} of the Nekrasov functions with $\epsilon_1$ playing role of  Planck's constant.
One may also consider a two-parametric deformation, where in this case one adds to the quantum integrable system its Whitham deformation controlled by the second deformation parameter. The latter is done within the framework of the AGT correspondence
\cite{AGT,Wyl,MM1'2009} and the most effective tools here are matrix models \cite{DV,IMO,EM'2009,EM'2010,SW'2009,MMSh'2009,MMSh'2010}. Note that at the level of AGT
with one of the deformation parameters set zero (so corresponding to a quantum integrable system \cite{NS,MM2'2009,MM3'2009,MT,MMM}), the $p-q$ duality
has quite an unexpected avatar: it is the spectral duality that describes a duality between the $SL(N)$ spin chain and the Gaudin model
with $N+2$ marked points (these two systems describe the two sides of the AGT correspondence) \cite{MMZZ13,MMRZZ13}. The spectral duality is lifted to $5d$ where it describes a duality of the $SL(N)$ spin chain on $M+2$ sites and the $SL(M)$ spin chain on $N+2$ sites \cite{MMRZZ34,GK2013}. Moreover, one can consider the full two-parametric AGT at this $5d$ level \cite{AHKSh,MMZ} where the spectral duality has a clear brane-picture interpretation.

However, an exact correspondence between  $p-q$ duality and  spectral duality has yet to be established. Moreover,
in contrast to the above cases, the probable generalizations of  spectral duality to $6d$ systems are not known even at the classical level, while the $p-q$ duality corresponding to $6d$ systems is known, as  explained above, and is described by the self-duality of the double-elliptic systems.

In this paper we continue the recent study \cite{AMMZ} of the double elliptic systems at the classical level. Our goal is to derive the theta-constant equations for the period matrix of the Seiberg-Witten curve of the double-elliptic integrable system. The main tool is the involutivity condition for the $N$-particle Hamiltonians constructed in \cite{BMMM'2000,MM,Braden2003} with respect to the Poisson bracket introduced in \cite{AMMZ}. Initially the Hamiltonians for the integrable systems under consideration were defined in \cite{BMMM'2000,MM} in the form of the ratios of theta-functions on Abelian varieties associated with the Jacobians of the corresponding Seiberg-Witten curves (see below). The hypothesis of \cite{MM} was that the Hamiltonians are Poisson commuting with respect to the Seiberg-Witten symplectic structure. Following this idea a new approach, which deals with arbitrary Riemann surfaces instead of the Seiberg-Witten curves was proposed in \cite{AMMZ}. Within this approach the concept of the Poisson bracket in terms of the coordinates on the Jacobian and the elements of the period matrix was introduced. An advantage of this approach is that it could lead to integrable systems not necessarily associated with the Seiberg-Witten curves. Indeed,
it was demonstrated in \cite{AMMZ} that the Poisson commutativity of the three-particle Hamiltonians is related just to some new theta-function identities of genus two, not making uses of  the Seiberg-Witten curve. In this paper, we extend the approach to the case of $N$-particle systems with $N>3$. When $N=4$ we describe evidence that the four-particle Hamiltonians are still in involution for an arbitrary Riemann surface of genus \eqs{g=3}.   However, in the general case when \eqs{N>4} the involutivity condition places restrictions on the period matrix and we find these satisfied for the special period matrices corresponding to the Seiberg-Witten curve of the double-elliptic integrable system, in accordance with the original expectation in \cite{MM}. Thus, the involutivity condition can be used to determine the dependence of the period matrix on the Seiberg-Witten flat moduli, providing an independent method for calculating the Seiberg-Witten prepotentials (including their instanton corrections).

In Section \ref{sec:Setting} we recall the definition of the $N$-particle Hamiltonians \cite{BMMM'2000,MM,Braden2003} and introduce the involutivity condition with respect to the Poisson bracket from \cite{AMMZ}. In Section \ref{sec:linProb} the involutivity condition is reformulated as a linear problem in terms of the vectors in a linear space of weight $3$ theta functions. In Section \ref{sec:example} we prove the theta-function identities that lie behind the Poisson commutativity of the three-particle Hamiltonians. In Section \ref{sec:SWrep} the method for calculating the Seiberg-Witten prepotentials of the double-elliptic integrable systems is presented. To illustrate the method we derive the Seiberg-Witten perturbative prepotential in the case of $N=5$.

\section{The Setting}
\label{sec:Setting}

Motivated by considerations of duality and the known rational, trigonometric and elliptic dependencies of the Calogero-Moser system the work of \cite{BMMM'2000} sought to construct integrable systems with compact momentum dependence, extending the rational and hyperbolic dependence of momentum of the Calogero-Moser  and  Ruijsenaars-Schneider systems respectively.  A 2-particle system that reproduced the Ruijsenaars-Schneider  (and so also the Calogero-Moser) system as a limit was constructed in which the
reduced momentum had elliptic dependence. That momentum dependence led to this class of models being called `double-elliptic', though actually the momenta posited in the paper more generally took values in some abelian variety $\mathfrak{A}$ (possibly a Jacobian) and the Hamiltonians were argued to be ratios of
theta functions, $\theta_a({\bf{z}}\,|\,\Omega)/\theta_b({\bf{z}}\,|\,\Omega)$ where $\bf{z}$ is the
momenta and $\Omega$ the period matrix of  $\mathfrak{A}$ . The relevant theta functions were introduced by analogy to those of the $N$-particle elliptic Calogero-Moser system. There  the genus $N$ spectral curve $\mathcal{C}$  is an $N$-fold covering of an elliptic curve $\mathcal{E}$  and the Jacobian $\Jac(\mathcal{C})$ is isogenous to an Abelian variety of the form $\Jac(\mathcal{E})\oplus \mathfrak{A}$. If $T$ is  the period matrix of $\mathcal{C}$  the general theory of coverings enables us to write\footnote{Here the Riemann $\theta$-function with
characteristics $\boldsymbol{a},\boldsymbol{b}\in\mathbb{Q}$  and $g\times g$ period matrix $T$ is
$$\theta\left[\begin{matrix}\boldsymbol{a}
\\ \boldsymbol{b}\end{matrix}\right](\boldsymbol{\hat z}\,|\,T)=
\sum_{\boldsymbol{n}\in\mathbb{Z}^g}\mathrm{exp}
\left\{\imath\pi(\boldsymbol{n}+\boldsymbol{a})^T T
            (\boldsymbol{n}+\boldsymbol{a})
+2\imath\pi
(\boldsymbol{n}+\boldsymbol{a})^T(\boldsymbol{\hat z}+\boldsymbol{b})
\right\}.
$$}
\begin{equation}
\label{thetadecomp}
\Theta(\mathbf{\hat z}| T)
=\sum_{\vec{\alpha}\in\mathbb{Z}^{N-1}/N\mathbb{Z}^{N-1}}
\theta\bsq{\at{-\frac1N\sum_{j=1}^{N}\alpha_{j}}{0}}\brc{z|\,N\tau}\,
\theta\bsq{\at{\frac{\vec{\alpha}}N}{\textbf{0}}}\brc{{\mathbf{z}}|\,\Omega}.
\end{equation}
where $(z, {\mathbf{z}})\sp{T}=M\mathbf{\hat z}\sp{T}$ for an appropriate $M$ separating out the centre of mass mode $z$ and $MTM\sp{T}=\begin{pmatrix}N\tau&0\\0&\Omega\end{pmatrix}$. Such decompositions are far from unique, for example the identity (here $e(x):=e\sp{2 i\pi x}$)
\begin{equation}
\label{eq:trans}
{\theta\bsq{\at{\textbf{0}}{\textbf{b}/l}}(\textbf{z}|l^{-1}\Omega)= \sum_{0\leq a_i<l} \rme\brc{\dfrac{\textbf{a}\cdot\textbf{b}}{l}}\theta\bsq{\at{\textbf{a}/l}{\textbf{0}}}(l\textbf{z}|l\Omega),\quad 0\leq b_i<l}
\end{equation}
with inverse
\begin{equation}
\theta\bsq{\at{\textbf{a}/l}{\textbf{0}}}(l\textbf{z}|l\Omega)= \dfrac1{l^g}\sum_{0\leq b_i<l} \rme\brc{-\dfrac{\textbf{a}\cdot\textbf{b}}{l}} \theta\bsq{\at{\textbf{0}}{\textbf{b}/l}}(\textbf{z}|l^{-1}\Omega),\quad 0\leq a_i<l
\end{equation}
lead to alternative expressions and we write these simply as
\begin{equation}
\Theta(\mathbf{\hat z}| T)
=\sum_a \theta_a\brc{z|\,N\tau}\, \theta_a
\end{equation}
specifying $\theta_a:=\theta_a({\mathbf{z}}|\,\Omega)$ as required. Although  \cite{BMMM'2000}  suggested the form of the Hamiltonians of the dual system, the exact nature of the $\theta_a$  was left unspecified (for $N>2$) and neither their Poisson commutativity nor their geometric setting was addressed at that time. The Seiberg-Witten picture tells us that $\Omega$ depends on the conjugate coordinates, but the dependence is left unspecified.

Before turning to the issue of Poisson commutativity we recall that Braden and Hollowood have given a geometric setting for such Hamiltonians \cite{Braden2003}. There the spectral curve is of genus $N+1$ and lies in a $(1,N)$-polarized abelian surface and is given by an equation of the form
\begin{equation}
0=\sum_{a=0}\sp{N-1}\Theta\left[\begin{matrix} 0&\frac{a}{N}\\ 0&0\end{matrix}\right]( z_1,z_2\,|\, \Gamma)\, \theta_a, \qquad
\theta_a=\sum_{{\vec{\alpha}\in\mathbb{Z}^{N-1}/N\mathbb{Z}^{N-1}} \atop {a+\sum_j \alpha_j\in N\mathbb{Z}  }}
\theta\bsq{\at{\frac{\vec{\alpha}}N}{\textbf{0}}}\brc{{\mathbf{z}}|\,\Omega}.
\end{equation}
Here $\Theta$ is the theta function of the abelian surface and
$\theta_a$  are of the form (\ref{thetadecomp}). Degenerations of this give precisely the Ruijsenaars and elliptic Calogero-Moser systems.

Regarding the Poisson commutativity of the ratios of theta functions (Hamiltonians) associated with the dual models of \cite{BMMM'2000},  significant evidence for this  was amassed in \cite{MM} using a perturbative (in instanton number) expansion.
Here it was observed that resulting equations for commutativity gave relations between the various terms of the instanton expansion of the Seiberg-Witten prepotential associated with $\mathcal{C}$. 
(A proof of the  Poisson commutativity in the case of the elliptic Calogero-Moser system was later provided in \cite{Marshakov:1999bw} .) Following \cite{BM01} we have that (for all $a,b,c,d$)
\begin{equation}
\label{posscom}
0=\left\{ {\theta_a\over\theta_b},
{\theta_{c}\over\theta_{d}}\right\}
\Longleftrightarrow
0=\sum_{i=1}\sp{N}
\left| \begin{array}{ccc}
\theta_{a}&\dfrac{\partial\theta_{a}}{\partial \hat z_{i}}&\dfrac{\partial\theta_{a}}{\partial \hat a\sp{i}}
\\
\theta_{b}&\dfrac{\partial\theta_{b}}{\partial \hat z_{i}} &\dfrac{\partial\theta_{b}}{\partial \hat a\sp{i}}
\\
\theta_{c}&\dfrac{\partial\theta_{c}}{\partial \hat z_{i}}&\dfrac{\partial\theta_{c}}{\partial \hat a\sp{i}}
\end{array}\right|
=
\sum_{r=1}\sp{N-1}
\left| \begin{array}{ccc}
\theta_{a}&\dfrac{\partial\theta_{a}}{\partial z_{r}}&\dfrac{\partial\theta_{a}}{\partial a\sp{r}}
\\
\theta_{b}&\dfrac{\partial\theta_{b}}{\partial  z_{r}} &\dfrac{\partial\theta_{b}}{\partial  a\sp{r}}
\\
\theta_{c}&\dfrac{\partial\theta_{c}}{\partial z_{r}}&\dfrac{\partial\theta_{c}}{\partial a\sp{r}}
\end{array}\right|
\end{equation}
where the action $\hat a\sp{i}$ is conjugate to the variable $\hat z_i$ and in obtaining the final equality we
express the simplectic form as
\[ \omega=\sum_{i=1}\sp{N}d \hat z_i \wedge d\hat a\sp{i}= dz\wedge d\tau +\sum_{r=1}\sp{N-1}
dz_r\wedge da\sp{r}
\]
noting that $\theta_a$ is independent of the centre of mass. Geometrically the Seiberg-Witten curve is the fibre over the moduli space of actions: this means the dependence of the theta functions on the action $\hat a\sp{i}$ is via the period matrix $T$ of $\mathcal{C}$ and hence $\Omega$.  Upon using the heat equation,
that $(\tau, {\mathbf{a}})\sp{T}=M\sp{-1\, T}\mathbf{\hat a}\sp{T}$, and the prepotential $\mathcal{F}$ we have that
\[ \dfrac{\partial\theta_{a}}{\partial  a\sp{r}}= \dfrac{\partial\theta_{a}}{\partial \Omega_{st}}
 \dfrac{\partial  \Omega_{st}}{\partial a\sp{r}}= \frac1{4i\pi} \dfrac{\partial\theta_{a}}{\partial  z_{r}z_s}\dfrac{\partial  \Omega_{st}}{\partial a\sp{r}}= \frac1{4i\pi} \dfrac{\partial\theta_{a}}{\partial  z_{r}z_s}
 \, M_{ri}M_{sj}M_{tk}\, \dfrac{\partial  T_{ij} }{\partial \hat a\sp{k}}
 =
 \frac1{4i\pi} \dfrac{\partial\theta_{a}}{\partial  z_{r}\partial  z_s}
 \, M_{ri}M_{sj}M_{tk}\,
 \dfrac{\partial\sp3  \mathcal{F} }{\partial \hat a\sp{i}\partial \hat a\sp{j}\partial \hat a\sp{k}};
\]
thus the Poisson-commutativity reduces to showing that
\begin{equation}
\label{pcomm}
0=
\sum_{r,s,t=1}\sp{N-1}P_{rst} H_{rst}\sp{abc},
\end{equation}
where $P_{rst}$ is totally symmetric (for this example $P_{rst}= M_{ri}M_{sj}M_{tk}\,
 {\partial\sp3  \mathcal{F} }/{\partial \hat a\sp{i}\partial \hat a\sp{j}\partial \hat a\sp{k}}  $) and
\begin{equation}
\label{eq:Hdet}
H_{rst}\sp{abc}:=
\left| \begin{array}{ccc}
\theta_{a}&\dfrac{\partial\theta_{a}}{\partial z_{r}}&\dfrac{\partial\theta_{a}}{\partial  z_{s}\partial  z_t}\\
\theta_{b}&\dfrac{\partial\theta_{b}}{\partial  z_{r}} &\dfrac{\partial\theta_{b}}{\partial  z_{s}\partial  z_t}\\
\theta_{c}&\dfrac{\partial\theta_{c}}{\partial z_{r}}&\dfrac{\partial\theta_{c}}{\partial  z_{s}\partial  z_t}\end{array}\right| .
\end{equation}
As noted above, this Poisson commutativity has been established for systems arising as degenerations of the elliptic Calogero-Moser system.

In the works just described we have utilised properties of the underlying spectral curve $\mathcal{C}$ to construct examples of double-elliptic systems. The recent paper \cite{AMMZ} goes beyond this. Let us henceforth assume Hamiltonians of the form ($ a=1,\dots,N-1$)
\begin{equation}
\label{eq:NHam}
{H_a\brc{\textbf{z}\,|\,\Omega}=\frac{\ds\theta\bsq{\at{0\;\dots\;0}{\frac{a}N\dots\frac{a}N}} \brc{\textbf{z}\,|\,\Omega}}{\theta\brc{\textbf{z}\,|\,\Omega}}
:=\frac{\theta_a \brc{\textbf{z}\,|\,\Omega}}{\theta_0\brc{\textbf{z}\,|\,\Omega}}  ,  }
\end{equation}
where $\Omega$ is $(N-1)\times (N-1)$ period matrix. (These are simply related to the Braden-Hollowood Hamiltonians via (\ref{eq:trans})\footnote{The Hamiltonians
$\displaystyle{\wt H_j=\frac{A_{0j}}{A_{00}} }$, $\displaystyle{A_{0j}=\sum_{\bfi{i_k}_j} \theta\bsq{\at{\frac{i_1}N\dots\frac{i_{N-1}}N}{0\,\,\dots\;\;\;\;0}} \brc{{\bf z}\,|\,\Omega}},
$
where \eqs{i_k=0,\dots,N-1} and the elements from \eqs{\bfi{i_k}_j} satisfy
${j+\sum_{k=1}^{N-1}i_k\in N\cdot\mathbb{Z}}$ may be expressed as
$${H_i\brc{{\textbf{z}}\,|\,\Omega}=\frac{\ds\theta\bsq{\at{0\;\dots\;0}{\frac{i}N\dots\frac{i}N}} \brc{{\bf z}/N\,|\,\Omega/N^2}}{\theta\brc{{\bf z}/N\,|\,\Omega/N^2}} =\frac{\ds\sum_{j=0}^{N-1}\rme\brc{-\frac{i\,j}N}A_{0j}}{\ds\sum_{j=0}^{N-1}A_{0j}},\quad i=1,\dots, N-1.}
$$
}.) Assuming only the Jacobi identity \cite{AMMZ}  sought
solutions to (\ref{pcomm}) for genus $2$ theta functions  ($N=3$). They discovered that $P_{rst}$
were totally symmetric and their solutions were expressed in terms of theta-function identities.
We reformulate and extend this as follows. Suppose one has a family of abelian varieties for which there is a symplectic structure on the total space with respect to which the abelian varieties are Lagrangian: we have in the above coordinates
\begin{equation}
\label{eq:PS0}
\{z_r,\Omega_{st}\}=P_{rst}(a),\quad
\{z_r,z_s\}=0,\quad \{\Omega_{rs},\Omega_{tu}\}=0,
\end{equation}
where the period matrix  $\Omega=\Omega(a)$ are some special functions of the Seiberg-Witten flat moduli $a$. We know from the work of Donagi and Markman \cite{DM96} that (assuming holomorphicity)  the differential of the period map at each point is the contraction of a cubic, and so $P_{rst}$ is totally symmetric. (This was a result of \cite{AMMZ}  that in the light of \cite{DM96}  we assume from the outset.)
What can be said about the solutions to (\ref{pcomm}) for arbitrary $g=N-1$? We will show that for \eqs{g=2} (\ref{pcomm}) holds for an arbitrary symmetric \eqs{ g\times g} period matrix $\Omega$ and are actually
theta-function identities. In the case when \eqs{g\geqslant4} the relations (\ref{pcomm}) define some special constraints on the elements of the period matrix $\Omega$.  We will show that these constraints describe the dependence of $\Omega$ on the flat moduli of the corresponding Seiberg-Witten curves. Thus, in the general case, the Poisson commutativity of the Hamiltonians (\ref{eq:NHam}) holds only for some special class of  period matrices, as  suggested in \cite{BMMM'2000,MM}.

\section{Poisson commutativity as a linear problem}
\label{sec:linProb}
The relations (\ref{pcomm}) are strongly connected with weight $3$ theta functions and their corresponding linear spaces  \cite{Mumford}.
Recall that an entire function \eqs{f\brc{\textbf{z}\,|\,\Omega}=f\brc{\textbf{z}}} on \eqs{\mathbb{C}^g} with fixed symmetric period matrix $\Omega$ is called  a theta function of weight $\lambda\in \mathbb{N}$ and characteristic $\bsq{\at{\boldsymbol{\delta}}{\boldsymbol{\epsilon}}}$, if
\begin{equation}
f(\mathbf{z}+\mathbf{p}\Omega+\mathbf{q})=e\left( -\frac{\lambda}2 \mathbf{p}\Omega\mathbf{p}-
\lambda \mathbf{p}\cdot\mathbf{z} + \boldsymbol{\delta}\cdot \mathbf{q}  -
\boldsymbol{\epsilon}\cdot \mathbf{p})
\right) f(\mathbf{z})
\end{equation}
for all $\mathbf{p},\mathbf{q}\in\mathbb{Z}^g$. Such functions form a linear space
$\Theta^{\Omega}_{\lambda}\bsq{\at{\boldsymbol{\delta}}{\boldsymbol{\epsilon}}}$
 of dimension \eqs{\lambda^g} with standard bases \cite{Fay1973}:
 \begin{align}
\label{eq:basis1}
(1)&\qq 
\theta
\left[\begin{matrix} \frac{\boldsymbol{\delta}+\boldsymbol{\rho} } {\lambda}\\ \boldsymbol{\epsilon}\end{matrix}\right]
\brc{\lambda\textbf{z}\,|\,\lambda\Omega},\quad 0\leqslant \rho_i<\lambda,\\
\label{eq:basis2}
(2)&\qq  
\theta
\left[\begin{matrix} \boldsymbol{\delta}\\ \frac{\boldsymbol{\epsilon}+\boldsymbol{\rho} }{\lambda} \end{matrix}\right]
\brc{\textbf{z}\,|\,\lambda^{-1}\Omega},\quad 0\leqslant \rho_i<\lambda.
\end{align}
It is convenient to define the general
\begin{equation}
\label{eq:cDet}{H^{\textbf{abc}}_{rst}\brc{\textbf{z}\,|\,\Omega}=\left|\matr{ccc}
{
\theta \bsq{\at{\textbf{a}}{\textbf{a}^{\prime} }} \brc{\textbf{z}\,|\,\Omega} &
\partial_{z_r}\theta \bsq{\at{\textbf{a}}{\textbf{a}^{\prime} }} \brc{\textbf{z}\,|\,\Omega}& \partial_{z_s}\partial_{z_t}\theta \bsq{\at{\textbf{a}}{\textbf{a}^{\prime} }}\brc{\textbf{z}\,|\,\Omega}\\
\\
\theta\bsq{\at{\textbf{b}}{\textbf{b}^{\prime} }}\brc{\textbf{z}\,|\,\Omega} &\partial_{z_r}\theta\bsq{\at{\textbf{b}}{\textbf{b}^{\prime} }}\brc{\textbf{z}\,|\,\Omega}& \partial_{z_s}\partial_{z_t}\theta\bsq{\at{\textbf{b}}{\textbf{b}^{\prime} }}\brc{\textbf{z}\,|\,\Omega}\\
\\
\theta\bsq{\at{\textbf{c}}{\textbf{c}^{\prime} }}\brc{\textbf{z}\,|\,\Omega}&
\partial_{z_r}\theta\bsq{\at{\textbf{c}}{\textbf{c}^{\prime} }}\brc{\textbf{z}\,|\,\Omega}& \partial_{z_s}\partial_{z_t}\theta\bsq{\at{\textbf{c}}{\textbf{c}^{\prime} }}\brc{\textbf{z}\,|\,\Omega}
}\right|,}
\end{equation}
where \eqs{ \textbf{a},\textbf{a}^{\prime} ,\textbf{b},\textbf{b}^{\prime} ,\textbf{c},\textbf{c}^{\prime}  \in\mathbb{Q}^g}.
With appropriate choices these will give  the determinants (\ref{eq:Hdet}). Now the symmetry of the coefficients $P_{rst}$ means that we
can work with the fully symmetric combinations
\begin{equation}
{H^{\textbf{abc}}_{{\{rst\}}}:=H^{\textbf{abc}}_{rst}+H^{ab}_{str}+H^{\textbf{abc}}_{trs},\quad r\leqslant s\leqslant t.}
\end{equation}
A simple calculation establishes that
\begin{equation}
H^{\textbf{abc}}_{\bfi{rst}}(\mathbf{z}+\mathbf{p}\Omega+\mathbf{q})=e\left( -\frac{3}2\, \mathbf{p}\Omega\mathbf{p}-  3\, \mathbf{p}\cdot\mathbf{z} +
\left[ \textbf{a}+\textbf{b}+\textbf{c}\right]\cdot \mathbf{q}  -
\left[ \textbf{a}^{\prime} +\textbf{b}^{\prime} +\textbf{c}^{\prime} \right]\cdot \mathbf{p})
\right)
H^{\textbf{abc}}_{\bfi{rst}}(\mathbf{z})
\end{equation}
and so
\begin{equation}
\label{Hwtheta}
H^{\textbf{abc}}_{\bfi{rst}}(\mathbf{z})\in \Theta^{\Omega}_{3}
\left[\begin{matrix}\textbf{a}+\textbf{b}+\textbf{c} \\  \textbf{a}^{\prime} +\textbf{b}^{\prime} +\textbf{c}^{\prime}.\end{matrix}\right]
\end{equation}
We emphasise that although none of the terms $H^{\textbf{abc}}_{rst}$ individually possess this property the symmetrised sum is a third order theta function. This result has several important consequences. First, let
$\{ f_{\vec{\alpha}}( \textbf{z}) \}$ be any basis for $\Theta^{\Omega}_{3}
\left[\begin{matrix}\textbf{a}+\textbf{b}+\textbf{c} \\  \textbf{a}^{\prime} +\textbf{b}^{\prime} +\textbf{c}^{\prime}.\end{matrix}\right]
$; then we have an expansion
\begin{equation}
\label{Hexpansion}
 H^{\textbf{abc}}_{\bfi{rst}}(\mathbf{z})=\sum_{ \vec{\alpha} } C_{\{rst\}}\sp{ \vec{\alpha}}\,  f_{\vec{\alpha}}( \textbf{z})
 \end{equation}
and (\ref{pcomm}) becomes for each $\vec{\alpha}$
\begin{equation}
\label{eq:g3ThetaId}
0=\sum_{r,s,t=1}\sp{g}P_{rst} \, C_{\{rst\}}\sp{ \vec{\alpha}}.
\end{equation}
This important relation entails several things. First it expresses that the $g\brc{g+1}\brc{g+2}/6$ vectors $C^{\vec{\alpha}}_{\bfi{rst}}$ (each one has \eqs{3^g} coordinates) are linearly dependent. Second, that we have a linear problem to determine the $P_{rst} $'s; and third, that the $P_{rst} $'s will be expressible in terms of the constants $C^{\vec{\alpha}}_{\bfi{rst}}$ if there is a nontrivial solution. Because the $C^{\vec{\alpha}}_{\bfi{rst}}$'s are given in terms of theta-constants, these are the theta-constant identities referred to earlier and generalise those obtained in \cite{AMMZ}. In the next section we shall illustrate this general setting for the Hamiltonians (\ref{eq:NHam}). For ease of description in what follows we will describe $C^{\vec{\alpha}}_{\bfi{rst}}$  as a $3^g \times g\brc{g+1}\brc{g+2}/6$ matrix.

\section{An Example}
\label{sec:example}
We shall now apply the above considerations to the Hamiltonians (\ref{eq:NHam}) and prove a conjecture raised in \cite{AMMZ}. The relevant characteristics are
\[ \bsq{\at{\textbf{a}}{\textbf{a}^{\prime} }}
=\left[ \begin{matrix} 0&\ldots&0\\ a/N&\ldots&a/N \end{matrix} \right], \quad
\bsq{\at{\textbf{b}}{\textbf{b}^{\prime} }}
=\left[ \begin{matrix} 0&\ldots&0\\ b/N&\ldots&b/N \end{matrix} \right], \quad
\bsq{\at{\textbf{c}}{\textbf{c}^{\prime} }}
=\left[ \begin{matrix} 0&\ldots&0\\ 0&\ldots&0 \end{matrix} \right],
\]
and we may obtain the expansion (\ref{Hexpansion}) as follows. The determinants \eqs{H^{\text{ab}}_{rst}} have the following Fourier decomposition:
\eqlb{eq:detFourier}{H^{\text{ab}}_{rst}=(2\pi\imath)^3\sum_{\textbf{n},\textbf{m},\textbf{l}\in\mathbb{Z}^g} \rme\brc{\frac12\textbf{n}^t\,\Omega\,\textbf{n}+\frac12\textbf{m}^t\,\Omega\,\textbf{m}+ \frac12\textbf{l}^t\,\Omega\,\textbf{l}+\brc{\textbf{n}+\textbf{m}+\textbf{l}}\cdot\textbf{z}}
\rme\brc{\frac{\textbf{a}\cdot\textbf{m}+\textbf{b}\cdot\textbf{l}}N}\abs{\textbf{n},\textbf{m},\textbf{l}}_{rst},}
where
\eq{\abs{\textbf{n},\textbf{m},\textbf{l}}_{rst}=\left|\matr{ccc}
{1 & n_r & n_s n_t\\
 1 & m_r & m_s m_t\\
 1 & l_r & l_s l_t
}\right|.}
The symmetrized functions $H^{\text{ab}}_{\bfi{rst}}$ then have an analogous expression in terms of
\eq{\abs{\textbf{n},\textbf{m},\textbf{l}}_{\bfi{rst}}= \abs{\textbf{n},\textbf{m},\textbf{l}}_{rst}+\abs{\textbf{n},\textbf{m},\textbf{l}}_{str} +\abs{\textbf{n},\textbf{m},\textbf{l}}_{trs}.}
That $H^{\textbf{ab}}_{\bfi{rst}}$  is a  theta functions of weight $\lambda=3$  is reflected by the
relation:
\eqlb{eq:detrel}{\forall\textbf{p}\in\mathbb{C}^g:\qq \abs{\textbf{n}+\textbf{p},\,\textbf{m}+\textbf{p},\,\textbf{l}+\textbf{p}}_{\bfi{rst}} =\abs{\textbf{n},\textbf{m},\textbf{l}}_{\bfi{rst}}.}
To obtain the decomposition in the basis (\ref{eq:basis1}) we change the summation variables in (\ref{eq:detFourier}) as follows
\eq{\left\{\matr{l}{
\textbf{n}\;\rightarrow3\textbf{k}+\vec{\alpha}-\textbf{m}-\textbf{l},\\
\textbf{m}\rightarrow\textbf{k}+\textbf{i},\\
\textbf{l}\;\,\,\rightarrow\textbf{k}+\textbf{j},}\right.
\qq
\textbf{i},\textbf{j},\textbf{k}\in\mathbb{Z}^g,
\quad\textrm{and}\quad
\vec{\alpha}\in\mathbb{Z}^g/3\mathbb{Z}^g.
}
Then, using the relation (\ref{eq:detrel}), we obtain
\eqlb{eq:Hdecomp}{H^{\text{ab}}_{\bfi{rst}}=\sum_{\vec{\alpha}\in\mathbb{Z}^g/3\mathbb{Z}^g}
C^{\vec{\alpha}}_{\bfi{rst}}\,
\,\theta\left[ \begin{matrix} &\vec{\alpha}/3&\\ (a+b)/N&\ldots&(a+b)/N \end{matrix} \right]
\brc{3\textbf{z}\,|\,3\Omega},}
where
\begin{equation}
\begin{split}
{C^{\vec{\alpha}}_{\{rst\}}=(2\pi\imath)^3
\rme\brc{-\frac{(a+b)\sum_k{\alpha_k}}{3N}}\,
\sum_{\textbf{i},\textbf{j}\in\mathbb{Z}^g}
\rme\brc{\brc{\textbf{i}+\frac{\textbf{j}-\vec{\alpha}}2}\Omega \brc{\textbf{i}+\frac{\textbf{j}-\vec{\alpha}}2}+
\brc{\frac{\textbf{j}}2-\frac{\vec{\alpha}}6}3\Omega
\brc{\frac{\textbf{j}}2-\frac{\vec{\alpha}}6}}\times
} \\
{\times
\rme\brc{\frac{\text{a}\sum_k i_k+\text{b}\sum_k j_k}N} \abs{\vec{\alpha}-\textbf{i}-\textbf{j},\,\textbf{i},\,\textbf{j}}_{\{rst\}}}
\end{split}
\end{equation}
or in terms of theta constants (with $\theta'_{r}\brc{0\,|\,\Omega}\equiv \left.\partial_{z_r}
\theta\brc{\textbf{z}\,|\,\Omega}\right|_{\textbf{z}=\textbf{0}}$)
\eqn{C^{\vec{\alpha}}_{\{rst\}}=2\,
\sum_{ \at{\vec{\beta}\in\mathbb{Z}^g/2\mathbb{Z}^g}{r,s,t}}
\left(9\theta'_{r}
\left[ \begin{matrix} &\frac{\vec{\beta}-\vec{\alpha}}2&\\ a/N&\ldots&a/N \end{matrix} \right]
\brc{0\,|\,2\Omega}\,
\theta''_{st}
	\bsq{\at{\frac{\vec{\beta}}2-\frac{\vec{\alpha}}6}{\brc{2\text{b}-\text{a}/N \ldots2\text{b}-\text{a}/N }} }
\brc{0\,|\,6\Omega}
\right.}
\eqlb{eq:thetacoord}{\left.-
\theta'''_{rst}
\left[ \begin{matrix} &\frac{\vec{\beta}-\vec{\alpha}}2&\\ a/N&\ldots&a/N \end{matrix} \right]
\brc{0\,|\,2\Omega}\,
\theta
\bsq{\at{\frac{\vec{\beta}}2-\frac{\vec{\alpha}}6}{\brc{2\text{b}-\text{a}/N \ldots2\text{b}-\text{a}/N }} }
\brc{0\,|\,6\Omega}\right).
}

Now when $N=3$  ($g=2$) we have just two Hamiltonians ($a=1$, $b=2$ or vica versa) and the functions \eqs{H^{\text{ab}}_{rst}} are invariant under the transformation
\eq{H^{\text{ab}}_{rst}\brc{\textbf{z}+\frac{\text{k}}3(1,1)\,|\,\Omega}= H^{\text{ab}}_{rst}\brc{\textbf{z}\,|\,\Omega}. }
In terms of the basis elements we have
\eq{\theta\bsq{\at{\vec{\alpha}/3}{\boldsymbol{\epsilon}}}\brc{3\brc{\textbf{z}+\frac{\text{k}}3(1,1) }\,|\,3\Omega}=
\rme\brc{\frac{ k \sum_l \alpha_l }3}
\theta\bsq{\at{\vec{\alpha}/3}{\boldsymbol{\epsilon}}}\brc{3\textbf{z}\,|\,3\Omega}.}
Hence the only nonzero coefficients in the decomposition (\ref{eq:Hdecomp}) are those \eqs{C^{\vec{\alpha}}_{\bfi{rst}}}  for which
\eqlb{eq:symRel}
{\rme\brc{\frac{ k \sum_l \alpha_l }3}=1
}
so leaving \eqs{3^{g-1}} possible nonzero coordinates in each column of \eqs{C^{\vec{\alpha}}_{\bfi{rst}}}.
Moreover,  the symmetry
\eq{H^{\text{ab}}_{rst}\brc{-\textbf{z}\,|\,\Omega}= H^{\text{ab}}_{rst}\brc{\textbf{z}\,|\,\Omega}}
reduces the independent terms further to only
${(3^{g-1}+1)/2}$ different nonzero rows in the matrix \eqs{C^{\vec{\alpha}}_{\bfi{rst}}}.
Now in the present setting we have
$g\brc{g+1}\brc{g+2}/{6}>{(3^{g-1}+1)}/{2}$ and so more variables $P_{\{rst\}}$ than equations, from which we deduce that for arbitrary  $\Omega$  equation (\ref{eq:g3ThetaId})
holds for nontrivial $P_{rst}$  and the corresponding $C^{\vec{\alpha}}_{\bfi{rst}}$ are linearly  dependent. This establishes  the theta-constant relations conjectured
in \cite{AMMZ}.

We remark in passing the above argument actually shows that for each pair of nonzero vectors \eqs{\textbf{a},\textbf{b}\in\mathbb{Z}^g/3\mathbb{Z}^g} with property \eqs{\textbf{a}+\textbf{b}\equiv\textbf{0}\mod 3} there exists a nontrivial set of quantities \eqs{P_{rst}} for which the corresponding Hamiltonians commute for
$g\le4$. This is because  $\textbf{z}\rightarrow\textbf{z}+{\textbf{a}}/3$ is still a symmetry of $H^{\textbf{ab}}_{rst}$, now leading to nonzero \eqs{C^{\vec{\alpha}}_{\bfi{rst}}}  when $\rme\brc{{\vec{\alpha}\cdot\textbf{a}}/3}=1$. Together with the restriction coming from parity, we have $\brc{g+1}\brc{g+2}/{6}>{(3^{g-1}+1)}/{2}$
for $g\le4$ and again there are corresponding theta-constant identities. When $\rme\brc{{\vec{\alpha}\cdot\textbf{a}}/3}\ne 1$ we find the identities $C^{\vec{\alpha}}_{\bfi{rst}}=0$.

\section{Theta-constant representation for the Seiberg-Witten curves}
\label{sec:SWrep}

In this section we shall use the constraint (\ref{pcomm}) to make various deductions about the prepotential.
This work will focus on the perturbative prepotential to establish the method, leaving the instanton corrections to a later work.

It is helpful to isolate the assumptions being made. First we are assuming that there exists an underlying Seiberg-Witten curve $\mathcal{C}$ with period matrix $T$ given by a prepotential $\mathcal{F}$.
Second, we shall assume that the curve $\mathcal{C}$ is of genus $N$ and covers an elliptic curve, and so is related to the elliptic Calogero-Moser family.  With a choice of $M$ given by
\eq{\forall i:\quad M_{1i}=M_{i1}=\dfrac{1}{N},\qq \forall i>1,j>1:\quad M_{ij}=-\dfrac{\delta_{ij}}N.}
we find
\eqlb{eq:OmegaSWN}{\Omega_{ij}=
\delta_{ij}\brc{\frac{g}{N}\tau-\sum_{k\neq i+1}T_{i+1,k}}+
(1-\delta_{ij})\brc{T_{i+1,j+1}-\frac1N\tau}
.}
In passing we note that at this stage one could have chosen  to use a different $M$ (for example \cite{BM01})
giving equivalent expansions, or chosen to have the Braden-Hollowood genus $N+1$ curve. Together these
assumptions provide us with an instanton expansion (for \eqs{i\neq j})
\eqlb{eq:SWTdec}{T_{ij}=\frac{\partial^2\mathcal{F}}{\partial\hat a_i\partial\hat a_j}=-\frac{1}{\pi\imath}\ln F^{(0)}_{ij}+\sum_{k\in\mathbb{N}}q^k\frac{\partial^2 F^{(k)}}{\partial\hat a_i\partial\hat a_j},\quad q\equiv \rme^{2\pi\imath\tau}.}
Here $F^{(0)}$ is the perturbative prepotential and the instanton corrections \eqs{F^{(k)}} to the Seiberg-Witten prepotential $\mathcal{F}$ are only known to low order. We may obtain new information about this expansion as follows. With
$P_{rst}= M_{ri}M_{sj}M_{tk}\, {\partial\sp3  \mathcal{F} }/{\partial \hat a\sp{i}\partial \hat a\sp{j}\partial \hat a\sp{k}}  $
 the constraints  (\ref{eq:g3ThetaId}) with (\ref{eq:thetacoord}) and the instanton expansions of these
 constrain the  $P_{rst}$'s. Let us write the perturbative expansions  (corresponding to the trigonometric limit \eqs{\textrm{Im}\,\tau\rightarrow+\infty}) as \eq{Q_{ij}\equiv\rme\brc{T_{ij}}= q_{ij}+ \sum_{k\in\mathbb{N}}q^k\,q^{(k)}_{ij},}
\eq{P_{ijk}=p_{ijk}+\sum_{l\in\mathbb{N}}q^{l}\,p^{(l)}_{ijk}.}
The constraints (\ref{eq:g3ThetaId}) become relations between the coefficients of these expansions. Before
illustrating this we note that although the dependence on the coordinates is not a priori known, we know  that in the case of the $GL\brc{N}$ systems under consideration (for example, from the Toda limits of \cite{MM})
the quantities $F^{(0)}_{ij}$ from (\ref{eq:SWTdec})  are just functions of $\brc{\hat a_i- \hat a_j}$, the difference of the Seiberg-Witten flat moduli. In general, the variables
${q_{ij}\equiv\brc{F^{(0)}_{ij}}^{-2}}$
may be represented by the following series:
\begin{equation}
\label{eq:qgenDec}
q_{ij}=\sum _{l=0} c_l \left(\hat a_i-\hat a_j\right)^{2l+2}.
\end{equation}

In the previous section we considered the case when $N=3$.  When $N=4$ we have $10$ vectors \eqs{C^{\vec{\alpha}}_{\bfi{rst}}} in a $27$ dimensional space. Consider the determinants \eqs{H^{ab}_{rst}} with $a=1$ and $b=3$. Then due to the symmetry
\eq{H^{ab}_{rst}\brc{-z\,|\,\Omega}=H^{ab}_{rst}\brc{z\,|\,\Omega}}
we have only $14$ different rows in the matrix consisting of vectors \eqs{C^{\vec{\alpha}}_{\bfi{rst}}}. The resulting $10\times14$ matrix has rank $9$ due to some theta-constant identities, thus giving the relations (\ref{pcomm}).

We therefore consider  the case $N=5$. Here we have $20$ vectors \eqs{C^{\vec{\alpha}}_{\bfi{rst}}} in an $81$ dimensional space.
Again, if we choose the determinants \eqs{H^{ab}_{rst}} with $a=1$ and $b=N-1=4$, we obtain $41$ different rows in the matrix consisting of vectors $C^{\vec{\alpha}}_{\bfi{rst}}$.
Then in order to satisfy the relations (\ref{pcomm}), the rank of this matrix must be at most equal to $19$. The latter gives at least $22$ equations on the elements of the period matrix \eqs{\Omega}. The number $22$ alone does not give useful information, as some of these equations may be equivalent. Now taking
the constraints (\ref{eq:g3ThetaId}) in the case of \eqs{N=5} and the perturbative expansions above, we obtain
the linear system
\eqlb{eq:LinSys}{L\bold{p}=0}
for the first nonzero order of the elements \eqs{p_{ijk}}:
\eq{i\neq j:\quad p_{ijj}=-p_{iij},\qquad
i\neq j\neq k:\quad p_{ijk}=0.}
Here
\begin{equation*}{
\bold{p}^{\textrm{T}}=
\left(\matr{cccccccccc}{
p_{111}& p_{112}& p_{113}& p_{114}& p_{222}& p_{223}& p_{224}& p_{333}& p_{334}& p_{444}}\right)}
\end{equation*}
(the remaining $p$'s appear at higher orders) and
\begin{equation*}
{L=}
{
\left(
\small
\begin{array}{cccccccccc}
 q_{13}-q_{23} & 0 & q_{13}-q_{23} & q_{13}-q_{23} & q_{12}-q_{23} & q_{12}-q_{23} & q_{12}-q_{23} & 0 & 0 & 0 \\
 q_{24}-q_{14} & q_{24}-q_{14} & 0 & q_{24}-q_{14} & 0 & q_{12}-q_{24} & 0 & q_{24}-q_{12} & q_{24}-q_{12} & 0 \\
 q_{25}-q_{15} & q_{25}-q_{15} & q_{25}-q_{15} & 0 & 0 & 0 & q_{12}-q_{25} & 0 & q_{12}-q_{25} & q_{25}-q_{12} \\
 0 & 0 & 0 & 0 & 0 & q_{35}-q_{45} & q_{34}-q_{45} & 0 & q_{34}-q_{35} & 0 \\
 0 & q_{14}-q_{34} & q_{13}-q_{34} & 0 & q_{34}-q_{14} & 0 & q_{34}-q_{14} & q_{34}-q_{13} & q_{34}-q_{13} & 0 \\
 0 & q_{15}-q_{35} & 0 & q_{13}-q_{35} & q_{35}-q_{15} & q_{35}-q_{15} & 0 & 0 & q_{13}-q_{35} & q_{35}-q_{13} \\
 0 & 0 & q_{25}-q_{45} & q_{24}-q_{45} & 0 & 0 & 0 & 0 & q_{24}-q_{25} & 0 \\
 0 & q_{25}-q_{35} & 0 & q_{23}-q_{35} & 0 & 0 & q_{23}-q_{25} & 0 & 0 & 0 \\
 0 & q_{24}-q_{34} & q_{23}-q_{34} & 0 & 0 & q_{23}-q_{24} & 0 & 0 & 0 & 0 \\
 0 & 0 & q_{15}-q_{45} & q_{14}-q_{45} & 0 & q_{15}-q_{45} & q_{14}-q_{45} & q_{45}-q_{15} & 0 & q_{45}-q_{14} \\
\end{array}
\right).}
\end{equation*}
Thus, the first constraint is of the form
\eqlb{eq:SWCons}{\det L=0.}

Now the constraint (\ref{eq:SWCons}) fixes the coefficients \eqs{c_l} in (\ref{eq:qgenDec}) for \eqs{l\geqslant4}. In particular, for \eqs{c_4} and \eqs{c_5} the constraint gives
\eq{c_4=\frac23\frac{c_1^4}{c_0^3}-\frac73\frac{c_1^2 c_2}{c_0^2}+ \frac23\frac{c_2^2}{c_0}+2\frac{c_1 c_3}{c_0},}
\eq{c_5=\frac{20}{33}\frac{c_1^5}{c_0^4}-\frac{49}{33}\frac{c_2 c_1^3}{c_0^3}+ \frac{14}{11}\frac{c_3 c_1^2}{c_0^2}-\frac{37}{33}\frac{c_2^2 c_1}{c_0^2}+ \frac{19}{11}\frac{c_2 c_3}{c_0}.}
Moreover, the constraint fixes the lowest term in the series (\ref{eq:qgenDec}) to be $1$ or \eqs{\left(\hat a_i-\hat a_j\right)^2} up to some constant factor. These recurrences are satisfied by
 the three sets of functions \eqs{q_{ij}}:
\begin{itemize}
\item{
corresponding to the $GL(N)$ elliptic Calogero system \eq{\label{tr1}q^{I}_{ij}=\brc{1-\frac{m^2}{\left(\hat a_i-\hat a_j\right)^2}}^{-1},}}
\item{
the $GL(N)$ elliptic Ruijsenaars model
\eq{\label{tr2}q^{II}_{ij}=\brc{1-\frac{m^2}{\sinh\left(\hat a_i-\hat a_j\right){}^2}}^{-1},}}
\item{the $GL(N)$ double elliptic system (where \eqs{\bar{\tau}} is the modulus of the second torus)
\eq{\label{tr3}q^{III}_{ij}=\brc{1-\frac{m^2}{\text{sn}\left(\left.\hat a_i-\hat a_j\right|\bar{\tau}\right){}^2}}^{-1}.}
}
\end{itemize}
Let us elaborate on the parameter count for the last of these (the former two being obtained as scaling limits of this). As we remarked above, there is an overall scaling of the functions $q_{ij}$ left undetermined by the constraint (\ref{eq:SWCons}). Further the scale $\sqrt{\mu}$ of the moduli $\hat a_i$, the mass $m$ and the period $\bar\tau$ are also parameters. Setting $q^{III}_{ij}={m^2}/{\mu}\brc{1-{m^2}/{\text{sn}\left(\left.\sqrt{\mu}\brc{\hat a_i-\hat a_j}\right|\bar{\tau}\right){}^2}}^{-1}$ scales $c_0=-1$ and the coefficients $c_{1,2,3}$ encode these
three parameters; the recursions for \eqs{c_l}  (\eqs{l\geqslant4}) then express the remaining coefficients
implicitly in terms of $\mu, m, \tau$.

The results just obtained suggest the following hypothesis. The constraints (\ref{eq:g3ThetaId}) are actually  theta-constant equations of the period matrix $T$ for the Seiberg-Witten curve associated with the $N$-particle double-elliptic integrable system for $N\geq 5$.
Thus, the equations can be used to determine the dependence of the period matrix on the Seiberg-Witten flat moduli, providing an independent method for calculating the Seiberg-Witten prepotentials (including the instanton corrections) of the $GL\brc{N}$ elliptic Calogero, Ruijsenaars and  double-elliptic systems.

\section*{Acknowledgements}
The work is partly supported by grant
NSh-1500.2014.2, by RFBR  grants 13-02-00457 (A.Mir.), 13-02-00478 (A.Mor.), 12-02-00594 (A.Z. and G.A.), by joint grants 13-02-91371-ST, 14-01-92691-Ind, by RFBR  grant 14-01-31353-mol\_a (G.A.),
by grant RSCF 14-50-00005 (A.Z.), by "Dynasty" fund (A.Z. and G.A.),
by the Brazil National Counsel of Scientific and
Technological Development (A.Mor.).

\bibliographystyle{unsrt}
\bibliography{references}

\end{document}